\documentclass[12pt]{article}
\usepackage[utf8]{inputenc}
\usepackage[left=1in,right=1in]{geometry}
\usepackage{authblk}
\usepackage[pdfstartview=FitH,pdfpagemode=None]{hyperref}
\usepackage{amsmath}
\usepackage{amssymb}
\usepackage{amsfonts}
\usepackage{tabularx}
\usepackage{braket}
\usepackage{cite}

\title{Krylov complexity has it all}
\author[1,2]{Wolfgang M\"uck}
\affil[1]{Dipartimento di Fisica ``Ettore Pancini'', Universit\`a degli Studi di Napoli Federico II \authorcr Via Cintia, 80126 Napoli, Italy}
\affil[2]{Istituto Nazionale di Fisica Nucleare, Sezione di Napoli \authorcr Via Cintia, 80126 Napoli, Italy}
\date{}

\begin{document}

%
\newcommand{\ie}{i.e.,\ }
\newcommand{\eg}{e.g.,\ }

\newcommand{\e}[1]{\operatorname{e}^{#1}}

\newcommand{\op}{\mathcal{O}}

\maketitle

\begin{abstract}
This paper establishes that Krylov complexity contains the entire information about the dynamics of a quantum operator, extending the list of equivalent quantities that can serve this purpose, such as the Lanczos coefficients, the return amplitude, and the spectral density. To demonstrate this equivalence, an explicit recursive algorithm is constructed to calculate Lanczos coefficients from the Taylor expansion of the Krylov complexity around $t=0$. Furthermore, the paper discusses the distinction between Krylov and spread complexity, clarifying why a similar recursive algorithm cannot exist for the latter without additional dynamical input. These results provide a ``proof of principle'' for using Krylov complexity as a complete characterization of operator evolution in quantum systems.
\end{abstract}

\paragraph{Introduction.}
The definition of Krylov complexity \cite{Parker:2018yvk} and the proposal of its use as a measure to characterize chaos in operator dynamics \cite{Parker:2018yvk, Barbon:2019wsy, Avdoshkin:2019trj, Caputa:2021sib, Rabinovici:2020ryf} has revived interest in the Recursion Method \cite{Viswanath-Mueller} and related Krylov subspace techniques based on the Lanczos algorithm \cite{Lanczos:1950}. Krylov complexity, in simple terms, is a time-dependent measure for the effective size (dimension) of the subspace of operators that a given observable explores during its time evolution. 
A similar quantity for quantum states is spread complexity, which measures the spreading of states in Hilbert space during their time evolution \cite{Balasubramanian:2022tpr, Alishahiha:2022anw, Erdmenger:2023wjg}. The excellent reviews \cite{Nandy:2024evd, Rabinovici:2025otw, Baiguera:2025dkc} provide an overview of these recent developments, with applications in fields that range from quantum computing to quantum gravity. 

It has been known for some time \cite{Viswanath-Mueller} that the dynamics of an operator $\op(t)$ can be represented in a variety of equivalent ways and that a number of quantities independently contain the entire information about this dynamics. This derives from the following chain of reasoning: In the Heisenberg picture, an operator $\op$ time-evolves as $\op(t) = \e{it\mathcal{L}} \op$, where $\mathcal{L}=[H,.]$ is the Liouvillian. Given an operator inner product, $(\op|\op')$, the Lanczos algorithm diagonalizes the set of operators $\mathcal{L}^n\op$, $n=0,1,2\ldots$, providing an ordered basis of operators $\op_n$, known as the Krylov basis, which satisfy the recurrence relations $\mathcal{L} \op_n = b_{n+1} \op_{n+1} + b_n \op_{n-1}$. The (non-negative) real numbers $b_{n}$ are known as Lanczos coefficients and represent the Liouvillian acting on the Krylov basis. Moreover, defining a chain of wavefunctions $\phi_n(t)\sim (\op_n|\op(t))$, the operator dynamics maps onto a hopping model on this (semi-infinite) chain. Furthermore, the recurrence relations define a set of real orthogonal polynomials with an associated orthogonality measure. Therefore, the following quantities contain, independently, the entire information about the dynamics of $\op(t)$: (i) the set of Lanczos coefficients, (ii) the return amplitude $\phi_0(t)$ or its Laplace transform, (iii) the spectral density associated with the orthogonal polynomials, and (iv) the moments of the spectral density, because they also appear as coefficients in the Taylor expansion of the return amplitude. 

In this paper I will establish that the above list can be extended by (v) Krylov complexity $K(t)$. In other words, knowing the Krylov complexity of an operator $\op(t)$ is equivalent to knowing its dynamics. For this purpose, I will explicitly construct an algorithm that recursively calculates the Lanczos coefficients from the coefficients of the Taylor expansion of $K(t)$ around $t=0$. This algorithm also can serve as a test whether a given (even) function $K(t)$ is compatible with a Krylov complexity. I will also discuss why this does not apply to spread complexity and provide a hint on what would be needed in this case.

\paragraph{Lanczos coefficients from Krylov complexity.}
The starting point is the hopping model on the one-dimensional chain associated with the time evolution of $\op(t)$ in the Krylov basis \cite{Viswanath-Mueller, Parker:2018yvk},\footnote{We follow the convention of \cite{Muck:2022xfc} absorbing factors of $i$ into the wavefunctions $\phi_n(t)$ such that all of them are real functions.} 
\begin{equation}
\label{alg:hopping}
	\partial_t \phi_n(t) = -b_{n+1} \phi_{n+1}(t) + b_n \phi_{n-1}(t)~, \qquad \phi_n(0) = \delta_{n,0}.
\end{equation}
Note that $b_n\geq 0$ is ensured by the Lanczos algorithm and $b_n=0$ for $n=N$ terminates it, leading to an $N$-dimensional Krylov space (and a chain of length $N$).
Define also
\begin{equation}
\label{alg:h.Delta}
	\Delta_n = b_n^2,\qquad h_n=\prod\limits_{k=1}^n \Delta_k,\qquad h_0=1~,
\end{equation}
which naturally appear in the \emph{monic} version of the Lanczos algorithm \cite{Muck:2022xfc} as the Lanczos coefficients ($\Delta_n$) and the norm of the $n$-th Krylov state ($h_n$). 
Then, it is clear that the solutions $\phi_n(t)$ of \eqref{alg:hopping} have Taylor expansions of the form
\begin{equation}
\label{alg:phi.expand}
	\phi_n(t) = \sqrt{h_n} \sum_{m=0}^\infty C_{n+m,m} \frac{t^{n+2m}}{(n+2m)!},
\end{equation} 
where the prefactor has been inserted for convenience. The hopping equation \eqref{alg:hopping} translates into the following recurrence formula for the matrix elements $C_{k,l}$,
\begin{equation}
\label{alg:C.recur}
	C_{k,l} = - \Delta_{k-l+1} C_{k,l-1} + C_{k-1,l}.
\end{equation}
The initial condition in \eqref{alg:hopping} implies $C_{0,0}=1$, and then \eqref{alg:C.recur} immediately gives $C_{k,0}=1\, \forall k$. The matrix $C_{k,l}$ is lower-triangular. An important point to remember is that the row $C_{p,\cdot}$ only contains the Lanczos coefficients $\Delta_q$ with $q\leq p$. 

Consider the Krylov complexity
\begin{equation}
\label{alg:Krylov.def}
	K(t) = \sum_{n=0}^\infty n \phi_n(t)^2,
\end{equation}
After substituting \eqref{alg:phi.expand} and performing a number of summation reorderings, one obtains
\begin{equation}
\label{alg:Krylov.expand}
	K(t) = \sum_{p=0}^\infty B_p \frac{t^{2p}}{(2p)!},
\end{equation}
with the coefficients\footnote{If, instead of the Krylov complexity, one considers the norm $\sum_{n=0}^\infty \phi_n(t)^2=1$, one obtains similar coefficients $A_p$, but without the factor $(p-n)$. It is a nice exercise to show that \eqref{alg:C.recur} ensures $A_p=\delta_{p,0}$.}  
\begin{equation}
\label{alg:B.p}
	B_p = \sum_{n=0}^p \sum_{k=0}^n \binom{2p}{p-n+2k} (p-n) h_{p-n} C_{p-n+k,k} C_{p-k,n-k}.
\end{equation}
The sum on the right-hand side of \eqref{alg:B.p} is symmetric under $k\leftrightarrow n-k$. In other words, if one considers the $(n,k)$-plane, on which the sums in \eqref{alg:B.p} are represented by the triangle $0\leq n\leq p$, $0\leq k \leq n$, the sum over the points above the line $n=2k$ is equal to the sum over the points below it. Therefore, \eqref{alg:B.p} can be expressed as
\begin{align}
\label{alg:B.p.rewrite}
	B_p =  \sum_{k=0}^{[p/2]} \Bigg[ & 2 \sum_{n=0}^{p-2k-1} \binom{2p}{p-n} (p-n-2k) h_{p-n-2k} C_{p-n-k,k} C_{p-k,n+k} \\
\notag & 
	- \binom{2p}{p} (p-2k) h_{p-2k} C_{p-k,k} C_{p-k,k} \Bigg].
\end{align}
The term on the second line compensates the double summation over the points on the line $n=2k$.

By simple inspection one finds that $B_p$ contains only the coefficients $\Delta_q$ with $q\leq p$, and that the coefficient $\Delta_p$ only appears in the terms with $k=0$. Thus, \eqref{alg:B.p.rewrite} can be split into 
\begin{equation}
\label{alg:B.zero.split}
	B_p = B_{p,1} + 2 \sum_{n=0}^{p-1} \binom{2p}{p-n} (p-n) h_{p-n} C_{p,n} - p\binom{2p}{p} h_p,
\end{equation} 
with
\begin{align}
\label{alg:B.p1}
	B_{p,1} =  \sum_{k=1}^{[p/2]} \Bigg[ & 2 \sum_{n=0}^{p-2k-1} \binom{2p}{p-n} (p-n-2k) h_{p-n-2k} C_{p-n-k,k} C_{p-k,n+k} \\
\notag & 
	- \binom{2p}{p} (p-2k) h_{p-2k} C_{p-k,k} C_{p-k,k} \Bigg].
\end{align}
The $k=0$ terms in \eqref{alg:B.zero.split} can be rewritten, by recursively applying \eqref{alg:C.recur} and using identities and sums for the binomial coefficients, as 
\begin{equation}
\label{alg:k.zero}
  2 \sum_{n=0}^{p-1} \binom{2p}{p-n} (p-n) h_{p-n} C_{p,n} - p\binom{2p}{p} h_p = 2 \binom{2p-2}{p-1} h_p + B_{p,2}
\end{equation} 
with 
\begin{equation}
\label{alg:B.p2}
	B_{p,2} = 4p \sum_{n=0}^{p-2} \binom{2p-2}{p-n-2} h_{p-n-1} C_{p-1,n+1},
\end{equation}
which contains only $\Delta_q$ with $q<p$. Thereby, the term with $h_p$ has been explicitly isolated in $B_p$.  

Using the above formulae, the algorithm given in Table~1 recursively calculates the Lanczos coefficients $\Delta_p$ from the Taylor series coefficients $B_p$ of Krylov complexity. Because the number of input parameters ($N$) is, in principle arbitrary, the Krylov complexity $K(t)$ contains the full information about the entire set of Lanczos coefficients. 

This does not mean that, given \emph{any} set of coefficients $B_p$ obtained from the Taylor series of some even function $F(t^2)$ with $F(0)=0$, one can find a set of Lanczos coefficients that produces the Krylov complexity $K(t)=F(t^2)$. In fact, the algorithm also serves as a test whether the input coefficients $B_p$ are compatible with a Krylov complexity. The caveat is that the construction involves the condition $\Delta_p\geq 0\, \forall p$. Therefore, as soon as the algorithm returns a single $\Delta_p<0$, it is clear that the input coefficients cannot represent a Krylov complexity. In the case, in which the algorithm stops at some $p=P$, because $\Delta_P=0$, one may be tempted to say that the Krylov space has dimension $P$, but then the remaining input coefficients $B_p$ for $p>P$ must uniquely depend on the Lanczos coefficients $\Delta_q$ with $q<P$, if they are to represent a Krylov complexity. Of course, this cannot be guaranteed either, and in most cases they simply will not. 
Finally, the algorithm is a proof of principle. Practical issues such as numerical stability are to be investigated.

\begin{table}[!t]
\begin{tabularx}{\textwidth}{|l|X|}
\hline
	\multicolumn{2}{|c|}{\textbf{Table 1: Krylov complexity to Lanczos coefficients algorithm}} \\
\hline
	Input & 
	coefficients $B_p$ ($p=1,2,\ldots,N$) in the Taylor series \eqref{alg:Krylov.expand}\\
\hline
	Start & 
	\hspace{1em}
	$h_0=1$, $C_{0,0}=1$ \\
\hline
	$p=1$ & \\[-0.5em]
	&
	\hspace{1em} $\displaystyle{h_1=\frac12 B_1, \qquad \Delta_1 =\frac{h_1}{h_0}}$, \qquad \text{Stop, if $\Delta_1\leq0$}.\\[1em]
	& 
	Prepare the data needed for the next step:\\
	&
	\hspace{1em} $C_{1,0}=1, \qquad C_{1,1} = -\Delta_1$.\\
\hline  
	$2\leq p\leq N$ &
	Calculate $B_{p,1}$ from \eqref{alg:B.p1} and $B_{p,2}$ from \eqref{alg:B.p2}.\\[0.5em]
	&
	\hspace{1em} $\displaystyle{h_p=\frac{B_p-B_{p,1}-B_{p,2}}{2 \binom{2p-2}{p-1}}, \qquad \Delta_p =\frac{h_p}{h_{p-1}}}$,  \qquad \text{Stop, if $\Delta_p\leq0$}.\\[1em]
	&
	Prepare the data needed for the next step: \\
	&
	\hspace{1em} $C_{p,0}=1, \qquad C_{p,q} = -\Delta_{p-q+1} C_{p,q-1} + C_{p-1,q}, \quad q=1,\ldots,p$.\\
\hline
	Result &
	Lanczos coefficients $\Delta_p$ ($p=1,2,\ldots,N$) \\
	& or incompatibility of the input, if the algorithm stops at $p<N$\\
\hline
\end{tabularx}
\end{table}

\paragraph{The case of spread complexity.}
A similar algorithm, which calculates the Lanczos coefficients $a_n$ and $b_n$ \cite{Viswanath-Mueller} of the Hamiltonian dynamics of a certain state from its spread complexity, cannot exist.
Spread complexity is defined by
\begin{equation}
\label{nogo:K}
	K_s(t) = \sum_{n=0}^\infty n |\phi_n(t)|^2,
\end{equation}
with the wave functions $\phi_n$ satisfying the chain of hopping equations\footnote{Other forms of the hopping equation, e.g.\ those in \cite{Erdmenger:2023wjg, Caputa:2024sux} differ from this by multiplying $\phi_n$ by suitable powers of $i$.}
\begin{equation}
\label{nogo:hopping}
	\partial_t \phi_n(t) = -b_{n+1} \phi_{n+1}(t) -i a_n \phi_n(t) + b_n \phi_{n-1}(t)~, \qquad \phi_n(0) = \delta_{n,0},
\end{equation}
where both, $a_n$ and $b_n$, are real, and one may take $b_n\geq 0$.  
Splitting the wave functions into their real and complex parts, $\phi_n = \varphi_n + i \psi_n$, \eqref{nogo:hopping} becomes a coupled system for real functions (with real coefficients) 
\begin{equation}
\label{nogo:hopsys}
\begin{split}
	\partial_t \varphi_n(t) &= -b_{n+1} \varphi_{n+1}(t) + a_n \psi_n(t) + b_n \varphi_{n-1}(t)~, \qquad \varphi_n(0) = \delta_{n,0},\\
	\partial_t \psi_n(t) &= -b_{n+1} \psi_{n+1}(t) - a_n \varphi_n(t) + b_n \psi_{n-1}(t)~, \qquad \psi_n(0) = 0.
\end{split}
\end{equation}
The solutions to these can be Taylor expanded as 
\begin{equation}
\label{nogo:Taylor}
\begin{split}
	\varphi_n(t) &= \sqrt{h_n} \sum_{m=0}^\infty C_{n+2m,2m} \frac{t^{n+2m}}{(n+2m)!},\\
	\psi_n(t) &= \sqrt{h_n} \sum_{m=0}^\infty C_{n+2m+1,2m+1} \frac{t^{n+2m+1}}{(n+2m+1)!},
\end{split}
\end{equation}  
where the coefficients have been arranged into a single matrix, which again turns out to be lower-triangular and can be generated by certain recurrence relations that follow from \eqref{nogo:hopsys}. In any case, as $K_s(t)$ in \eqref{nogo:K} depends on the norm $|\phi_n(t)|^2 = \varphi_n(t)^2 + \psi_n(t)^2$, its Taylor expansion only contains even powers of $t$.\footnote{This fact is immediately relevant for checking the holographic Krylov--momentum proposal \cite{Caputa:2024sux, Fan:2024iop, Rabinovici:2023yex, He:2024pox, Xu:2024gfm, Ambrosini:2024sre, Li:2025fqz, Aguilar-Gutierrez:2025mxf, Ambrosini:2025hvo, Nastase:2026lhz, Zoakos:2026obl, Fu:2025kkh, Jeong:2026iac, Fatemiabhari:2025cyy, Fatemiabhari:2025poq, Fatemiabhari:2025usn, Fatemiabhari:2026goj, Fatemiabhari:2026rob, Alfinito:2026vah, Chatzis:2026ekd}.} Therefore, spread complexity $K_s(t)$ offers exactly the same amount of input information as Krylov complexity $K(t)$. Since the algorithm in Table~1, if it does not fail, uniquely determines the Lanzcos coefficients $\Delta_n=b_n^2$ (with $a_n=0$), there cannot be an algorithm which determines both, $a_n$ and $b_n$, from the same type of input.

One may wonder what kind of additional input, together with spread complexity, would be needed such that the entire set of Lanczos coefficients can be recursively determined.\footnote{A shift $a_n\to a_n+c$ corresponds to an overall shift of energy, which does not change the complexity functions. Therefore, the coefficient $a_0$ can always be arbitrarily chosen.} For example, one may consider the second spread complexity, 
\begin{equation}
\label{nogo:K2}
	K_2(t) = \sum_{n=0}^\infty n^2 |\phi_n(t)|^2.
\end{equation}
Explicitly considering the first few Taylor coefficients of $K_s(t)$ and $K_2(t)$ one realizes that, at order $t^{2p}$ they both contain combinations of  $\Delta_q$ with $q\leq p$ and $(a_q-a_0)$ with $q<p$. More precisely, the $t^2$ coefficients of $K_s(t)$ and $K_2(t)$ coincide and are given by $2\Delta_1$, so that one can determine $\Delta_1$ (and arbitrarily choose $a_0$). From the $t^4$ coefficients one can calculate $\Delta_2$ and $(a_1-a_0)$, and so on. Therefore, a recursive algorithm to determine all $\Delta_p$ and $(a_p-a_0)$ seems possible, with the caveat that it terminates, if $\Delta_p\leq0$. The explicit construction of such an algorithm or some other compatibility test would be interesting and also contribute to our understanding of the holographic Krylov--momentum proposal.

\paragraph{Acknowledgements.}
I thank Ali Fatemiabhari and Carlos Nu\~nez for stimulating discussions.
Partial support by the INFN under the research initiative STEFI is gratefully acknowledged.

\providecommand{\href}[2]{#2}\begingroup\raggedright\endgroup

\end{document}